\documentclass[aps,prx,reprint,superscriptaddress,longbibliography]{revtex4-2}
\usepackage{amsmath,amssymb,graphicx, bm,xcolor,hyperref}
\usepackage[utf8]{inputenc}
\usepackage[T1]{fontenc}
\usepackage{braket}
\usepackage{booktabs}
\usepackage[normalem]{ulem}
\usepackage{siunitx}

\usepackage{xcolor}

\begin{document}

\title{Unified Flux Control Architecture for Fluxonium Qubits}

\author{Xianchuang Pan}
\thanks{These authors contributed equally to this work.}
\affiliation{S Lab, Quantum Science Center of Guangdong-Hong Kong-Macao Greater Bay Area, Shenzhen, China}

\author{Jiahui Wang}
\thanks{These authors contributed equally to this work.}
\affiliation{S Lab, Quantum Science Center of Guangdong-Hong Kong-Macao Greater Bay Area, Shenzhen, China}

\author{Tao Zhou}
\thanks{These authors contributed equally to this work.}
\affiliation{S Lab, Quantum Science Center of Guangdong-Hong Kong-Macao Greater Bay Area, Shenzhen, China}
\author{Yanbo Guo}
\author{Fei Wang}
\author{Ze Zhan}
\author{Liang Xiang}
\author{Zishuo Li}
\author{Lu Ma}
\author{Xizheng Ma}
\author{Huijuan Zhan}
\author{Tao Zhang}
\author{Kannan Lu}
\author{Xing Zhu}
\email{zhuxing@quantumsc.cn}
\author{Guicheng Gong}
\author{Chunqing Deng}
\email{dengchunqing@quantumsc.cn}
\author{Tenghui Wang}
\email{wangtenghui@quantumsc.cn}
\affiliation{S Lab, Quantum Science Center of Guangdong-Hong Kong-Macao Greater Bay Area, Shenzhen, China}

\date{\today}

\begin{abstract}
Control architectures that reduce hardware overhead while maintaining high-fidelity operations are essential for the continued scaling of superconducting quantum processors. Here we experimentally realize a unified control architecture for fluxonium qubits, in which both transverse ($XY$) and longitudinal ($Z$) operations are implemented through a single flux-control channel driven by a single arbitrary waveform generator channel. This architecture imposes competing requirements on the shared control channel, which must simultaneously support low-frequency flux transmission for reset operations while strongly attenuating broadband noise near the qubit transition frequency. We address this challenge through frequency-selective cryogenic filtering together with compensated waveform synthesis that corrects the pulse distortion introduced by the filtered control line. Experimentally, this approach preserves coherence times above 100~$\mu$s while enabling active reset with approximately 98\% fidelity and 20-ns single-qubit gates with fidelities exceeding 99.99\%. We further demonstrate FPGA-native instruction-level waveform synthesis based on reusable pulse primitives for unified flux control.
These results establish unified flux control as a scalable architecture for fluxonium qubits that reduces control hardware overhead while preserving high-fidelity operation.
\end{abstract}

\maketitle
\section{Introduction}

Superconducting quantum circuits are among the leading platforms for quantum information processing, with rapid advances in coherence~\cite{place2021new,ganjam2024surpassing,nguyen2019high,Somoroff2023,Wang2025}, gate fidelity~\cite{sung2021realization,negirneac2021,ding2023high,rower2024suppressing,zhan2026scalable}, and processor scale~\cite{arute2019quantum,Wu2021,kim2023evidence,van2024advanced,Jin2025}. As these systems continue to scale, however, the complexity of the control architecture itself increasingly becomes a limiting factor. In most superconducting platforms, universal qubit control relies on multiple physically distinct channels: transverse ($XY$) operations are implemented using microwave charge drive, while longitudinal ($Z$) control is realized through low-frequency flux bias~\cite{Koch2007,manucharyan2009fluxonium,krantz2019quantum}. This separation requires duplicated electronic sources, filtering stages, and calibration abstractions. Several approaches have attempted to reduce wiring complexity by integrating microwave and flux pathways within a shared on-chip structure~\cite{manenti2021full,yan2026characterizing}. However, these approaches still rely on multiple independent control sources operating across different spectral regimes. As a result, the underlying control paradigm remains fundamentally multi-channel, inherently limiting further architectural simplification.

Unlike superconducting qubits based on plasmon-like excitations in circuits such as the transmon, flux-like qubits such as fluxonium derive their qubit transitions from fluxon tunneling. In these systems, both transverse and longitudinal control can be implemented through the same inductive flux degree of freedom~\cite{Zhang2021Universal,Deng2015Observation, ma2023native,  rower2024suppressing}. Accordingly, transverse rotations are driven by modulated ac flux signals, while longitudinal operations are implemented through quasi-dc flux bias. Ultimately, universal single-qubit control can be collapsed into a single inductive channel. This unified control channel enables architectural simplification while remaining compatible with the operating regime in which fluxonium qubits have demonstrated long coherence times~\cite{nguyen2019high,Somoroff2023,Wang2025} and high-fidelity gate performance~\cite{ding2023high,rower2024suppressing,zhan2026scalable,Zhang2021Universal,Ficheux2021,bao2022fluxonium,huang2023quantum,ma2023native,wang2024efficient}.

However, consolidating universal control into a shared physical channel introduces nontrivial engineering challenges. The same line must simultaneously preserve strong low-frequency transmission for large flux excursions and active reset~\cite{mcewen2021removing,Reed2010fast,wang2024efficient}, while strongly attenuating broadband noise near the qubit transition frequency to protect coherence ~\cite{krinner2019engineering,Simbierowicz2024Inherent}. In practice, the required line response is difficult to realize: the channel must be nearly transparent near dc while providing spectrally flat attenuation around the qubit transition frequency (hundreds of MHz). Residual amplitude/phase ripple then distorts the microwave flux modulation and degrades $XY$-control fidelity. Approaches exploiting higher-order nonlinear processes have recently been explored to partially relax this filtering constraint by enabling control at subharmonic frequencies through filtered lines~\cite{xia2025fast,schirk2025subharmonic}. However, these schemes generally require higher-order nonlinear processes and introduce additional calibration complexity, including amplitude-dependent frequency shifts and reduced effective coupling strengths.

In this work, we experimentally realize a unified control architecture for fluxonium qubits through a joint design of cryogenic filtering and compensated waveform synthesis. We first show that frequency-selective cryogenic filtering preserves long coherence while maintaining the flux tuning range required for flux-controlled reset. We then introduce a bounded inverse-filtering framework that compensates waveform distortion, enabling high-contrast Rabi oscillations, active reset, and single-qubit gate fidelities above $99.99\%$ within a unified flux control channel. Building on this calibrated control layer, we further implement FPGA-native instruction-level waveform synthesis based on reusable pulse primitives, enabling complex control sequences to be dynamically generated within the same unified control framework. Together, these results establish flux control as a unified architecture for fluxonium qubits.

\section{Unified flux control Architecture}

\begin{figure*}[bt]
  \includegraphics{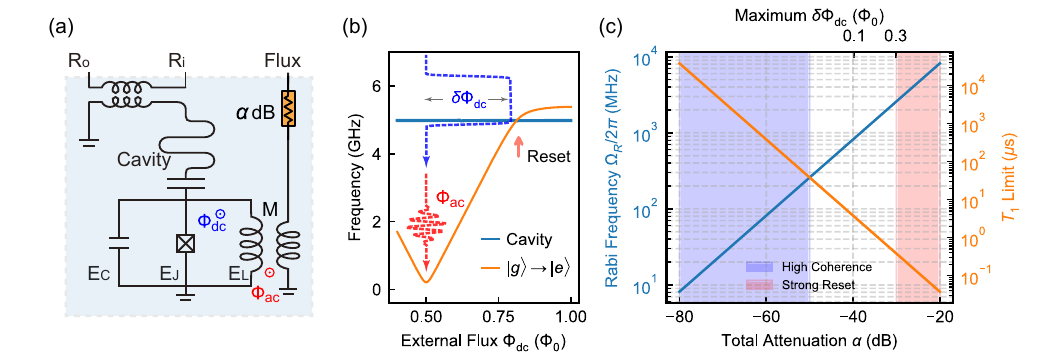}
  \caption{\textbf{Unified Flux Control Architecture}. 
    (a) Circuit schematic of a fluxonium qubit coupled to a single unified flux line. The control signals undergo a total attenuation of $\alpha$ before coupling to the qubit via a mutual inductance $M$. 
    (b) Energy spectrum of the fluxonium and the readout cavity as a function of the external static flux $\Phi_{\rm dc}$. The dashed arrows illustrate the multiplexed control protocol at the half-flux sweet spot ($\Phi_{\rm dc} = 0.5\Phi_0$): a microwave pulse $\Phi_{\rm ac}$ drives the transverse $|g\rangle \rightarrow |e\rangle$ transition for single-qubit gates, while a baseband pulse $\delta\Phi_{\rm dc}$ longitudinally tunes the qubit frequency toward the cavity for rapid initialization. 
    (c) Calculated Rabi frequency $\Omega_R/2\pi$ (blue, left axis) and the $T_1$ relaxation limit induced by room-temperature wideband noise (orange, right axis) versus the total line attenuation $\alpha$. The top axis indicates the corresponding maximum achievable DC flux excursion. The shaded areas denote the ``High Coherence'' regime (blue) and the ``Strong Reset'' regime (red), illustrating the strict necessity for frequency-selective filtering in the single-port design.}
  \label{fig1}
\end{figure*}

In this section, we present the unified control architecture for the fluxonium qubit. As depicted in Fig.~\ref{fig1}(a), unlike conventional architectures that require separate charge and flux lines, our design utilizes a single unified flux line to mediate all necessary control signals. The system is governed by the fluxonium Hamiltonian, 
\begin{equation}
    H_0 = 4E_C \hat{n}^2 + \frac{1}{2}E_L (\hat{\phi}- \phi_{\rm ext})^2 - E_J \cos(\hat{\phi}),
\end{equation}
where $E_C$, $E_L$, and $E_J$ are the charging, inductive, and Josephson energies, respectively, and $\phi_{\rm ext}=2\pi\Phi_{\rm ext}/\Phi_0$ is the reduced external magnetic flux. The control signals are synthesized by an arbitrary waveform generator (AWG) at room temperature, attenuated by a total factor of $\alpha$ inside the dilution refrigerator, and inductively coupled to the qubit through a mutual inductance $M$. Consequently, the total external flux can be decomposed into a static bias and time-dependent control components,
\begin{equation}
    \Phi_{\rm ext}(t)=\Phi_{\rm dc}+\delta\Phi_{\rm dc}(t)+\Phi_{\rm ac}(t).
\end{equation}

Fig.~\ref{fig1}(b) illustrates the fluxonium spectrum together with the corresponding unified control protocol. The qubit is typically biased at the half-flux sweet spot ($\Phi_{\rm dc}=0.5\Phi_0$), where first-order sensitivity to flux noise vanishes. At this operating point, the $|g\rangle \rightarrow |e\rangle$ transition frequency is strongly suppressed and typically lies in the 200--400 MHz range. As indicated by the dashed arrows in Fig.~\ref{fig1}(b), the shared flux-control line simultaneously supports universal control by applying both a microwave component $\Phi_{\rm ac}(t)$ for resonant $XY$ rotations and a baseband component $\delta\Phi_{\rm dc}(t)$ for dynamically shifting the qubit to higher frequencies during initialization. When the readout cavity frequency is designed below the maximum qubit transition frequency, dynamic flux tuning can further bring the qubit close to resonance with the cavity, enhancing qubit--cavity interaction and enabling accelerated reset. In this way, universal single-qubit control is implemented within a single flux control framework.

For the spectrum shown in Fig.~\ref{fig1}(b), we adopt device parameters of $E_J/h=4.5$~GHz, $E_C/h=1.1$~GHz, and $E_L/h=0.5$~GHz, together with a readout resonator frequency $f_r=4.98$~GHz. These values are consistent with experimentally realized fluxonium devices and reflect typical operating conditions in our setup. However, implementing universal control through a shared flux-control line introduces an intrinsic bandwidth-distortion tradeoff. From the linear expansion of the inductive energy~\cite{you2019circuit}, the effective Rabi drive strength $\Omega_R$ is directly proportional to the total line transmission factor $\alpha$:
\begin{equation}
    \hbar \Omega_R =
    \left|
    2\pi E_L \frac{M}{\Phi_0} \frac{\alpha}{Z_0}
    \right|
    V_0 |\langle 0|\hat{\phi}|1\rangle|,
\end{equation}
where $V_0$ is the room-temperature drive amplitude and $Z_0$ is the line impedance. At the same time, wideband noise from room-temperature electronics propagates through the same control line and induces qubit relaxation through environmental voltage fluctuations~\cite{SetePurcell2014}. Within a Fermi’s Golden Rule treatment, the corresponding relaxation rate $\Gamma_{1\rightarrow0}^{\rm line}=1/T_1^{\rm line}$ scales quadratically with the line transmission factor $\alpha$,
\begin{equation}
    \frac{1}{T_1^{\rm line}}
    =
    \frac{1}{\hbar^2}
    \left(
    2\pi E_L \frac{M}{\Phi_0} \frac{\alpha}{Z_0}
    \right)^2
    |\langle 0|\hat{\phi}|1\rangle|^2
    S_{VV}^{\rm awg}(\omega_{ge}),
\end{equation}
where $S_{VV}^{\rm awg}(\omega_{ge})$ denotes the effective voltage-noise spectral density at the AWG output. The detailed calculation is summarized in Appendix~\ref{mode}.

To quantify this tradeoff, we assume a mutual inductance of $M=2$~pH and a typical commercial AWG with active wideband noise characterized by a power spectral density of $-130$~dBm/Hz. Fixing the maximum AWG output amplitude at $V_0=0.5$~V, we evaluate the resulting coherence limit together with the achievable controllability for gate operations and initialization across different attenuation factors. The resulting tradeoff defines a central constraint of the unified architecture, as summarized in Fig.~\ref{fig1}(c). In the high-coherence regime (heavy attenuation, $\alpha \approx -80$ to $-50$~dB), the qubit maintains coherence times exceeding $10^2~\mu$s, but only weak drive amplitudes reach the device, resulting in slow gate operations. Conversely, in the strong-reset regime (light attenuation, $\alpha \approx -30$ to $-20$~dB with maximum $\delta\Phi_{\rm dc}/\Phi_0 \ge 0.3$), large flux excursions become accessible for rapid modulation and reset, but coherence is strongly degraded by broadband noise injection. These results highlight that realizing this unified architecture requires careful engineering of the attenuation chain, strictly necessitating the deployment of frequency-selective filtering to suppress $S_{VV}^{\rm awg}(\omega_{ge})$ while maintaining high signal transparency at baseband frequencies.

\section{Experimental demonstration of Unified Flux Control}

\begin{figure*}
  \includegraphics{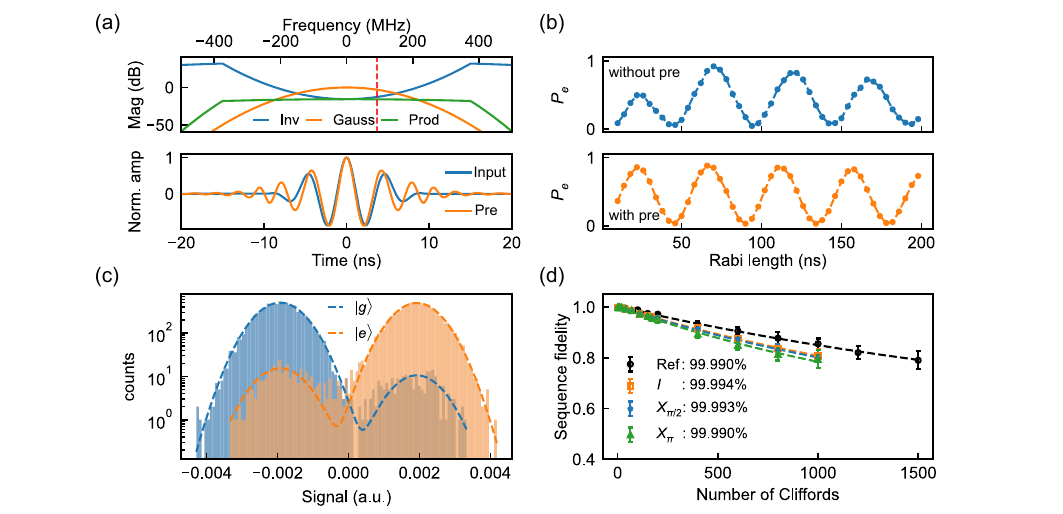}
  \caption{\textbf{Experimental Demonstration of Unified Flux Control.}
(a) Frequency- and time-domain illustration of the pre-distortion procedure. Top: the bounded inverse filter $H_{\rm inv}(f)$ (blue), the Gaussian low-pass transfer function $H_{\rm gauss}(f)$ (orange), and the resulting effective response $H_{\rm prod}(f)=H_{\rm gauss}(f)H_{\rm inv}(f)$ (green). Bottom: comparison between the target input waveform (blue) and the pre-distorted waveform generated by the AWG (orange). 
(b) Rabi oscillations measured without pre-distortion (top, blue) and with pre-distortion (bottom, orange). Pre-distortion restores high-contrast oscillations and enables nearly complete population transfer. 
(c) Histograms of the rotated readout signal together with Gaussian fits for the $|g\rangle$ (blue) and $|e\rangle$ (orange) state references, used to estimate the residual excited-state population after the reset protocol. 
(d) Standard randomized benchmarking and interleaved randomized benchmarking for 20-ns single-qubit gates. Black circles: reference RB; orange squares: interleaved $I$ gate; blue diamonds: interleaved $X_{\pi/2}$ gate; green triangles: interleaved $X_{\pi}$ gate. The extracted gate fidelities are all at or above $99.99\%$.}
  \label{fig2}
\end{figure*}

We now implement unified flux control using frequency-selective filtering to resolve the coherence-versus-control trade-off identified in the previous section. The detailed cryogenic setup is provided in Appendix~\ref{Setup}. For a fluxonium qubit with transition frequency $f_q = 208$~MHz, the control line incorporates a Gaussian low-pass filter with a 3-dB cutoff frequency of $f_c \approx 92$~MHz and transfer function $H_{\rm gauss}(f)=\exp[-(f\sigma)^2/2]$, providing approximately 18~dB attenuation at the qubit frequency. Combined with the 30~dB cryogenic attenuation and cable losses, the total attenuation exceeds 50~dB, strongly suppressing broadband noise from room-temperature electronics. Under these conditions, the qubit remains within the high-coherence regime, with measured coherence times of $T_1=110~\mu\text{s}$, $T_2^{\rm R}=128~\mu\text{s}$, and $T_2^{\rm echo}=133~\mu\text{s}$. A detailed coherence analysis is provided in Appendix~\ref{coherence}.

The same filtering, however, inevitably distorts short control pulses because of its finite bandwidth and associated phase dispersion. To compensate for this effect, we apply a pre-distortion procedure that reshapes the input waveform such that the combined response of the control line is approximately flat over the qubit operation band. Instead of implementing a direct reciprocal filter, which would diverge at high frequencies, we construct a bounded inverse filter that selectively compensates attenuation near the qubit frequency while remaining stable elsewhere. The inverse filter is defined as
\begin{equation}
    H_{\rm inv}(f) =
    \frac{H_{\rm qubit}}
    {\max[H_{\rm gauss}(f),10^{-G_{\rm max}/20}]}
    \times W(f),
\end{equation}
where $H_{\rm qubit}=H_{\rm gauss}(f_q)$ sets the normalization at the qubit frequency. The maximum gain is capped at $G_{\rm max}=50$~dB to avoid numerical divergence and AWG saturation, while the Gaussian window $W(f)=\exp[-(f\sigma_w)^2/2]$
suppresses compensation outside the target bandwidth with a 3-dB cutoff near 1~GHz.

The resulting compensation procedure is illustrated in Fig.~\ref{fig2}(a). The upper panel compares the bounded inverse filter $H_{\rm inv}(f)$ (blue), the Gaussian line response $H_{\rm gauss}(f)$ (orange), and the resulting effective transfer function $H_{\rm prod}(f)=H_{\rm gauss}(f)H_{\rm inv}(f)$ (green). The compensated response remains approximately flat around $f_q$ while preserving strong attenuation outside the operational band. In the time domain, shown in the lower panel of Fig.~\ref{fig2}(a), the pre-distorted waveform generated by the AWG contains additional high-frequency components that counteract the filtering-induced distortion, thereby recovering the intended pulse shape at the qubit. We first implement this compensation entirely in software, without additional hardware modifications.

The necessity and effectiveness of the pre-distortion procedure are verified experimentally through the Rabi measurements shown in Fig.~\ref{fig2}(b). Without pre-distortion, the oscillations exhibit strongly reduced contrast and fail to achieve complete population inversion, particularly for short pulses with broad spectral content. With pre-distortion enabled, high-contrast sinusoidal Rabi oscillations are recovered, confirming that accurate single-port control can be maintained despite the strong low-pass filtering. We next demonstrate the active-reset protocol enabled within the same unified control framework. Following the operating principle illustrated in Fig.~\ref{fig1}(b), a quasi-dc flux excursion of $\Delta \Phi_{\rm dc} \approx 0.298 \Phi_0$ is applied to bring the qubit close to resonance with the readout cavity, where it remains for approximately $5~\mu$s to enhance energy relaxation. Fig.~\ref{fig2}(c) shows the rotated readout histograms together with Gaussian fits for the $|g\rangle$ (blue) and $|e\rangle$ (orange) state references~\cite{bao2022fluxonium}. From the overlap analysis, we extract a residual excited-state population of approximately 2\%, corresponding to an initialization fidelity near 98\%. Although the present protocol is not optimized for speed, substantially faster initialization should be achievable with optimized flux trajectories and device parameters.

Finally, we benchmark single-qubit gate performance within the unified flux-control architecture. We employ standard cosine pulse envelopes and follow a calibration protocol identical to that described in Ref.~\cite{bao2022fluxonium}. Using a gate duration of $20$\,ns, we perform standard randomized benchmarking (RB) together with interleaved randomized benchmarking (iRB)~\cite{magesan2012efficient}, as shown in Fig.~\ref{fig2}(d). The black circles denote the reference RB sequence, while the orange squares, blue diamonds, and green triangles correspond to interleaved benchmarking of the $I$, $X_{\pi/2}$, and $X_{\pi}$ gates, respectively. The extracted fidelities for these representative single-qubit gates are all at or above $99.99\%$. Furthermore, as detailed in the Appendix~\ref{all_devices}, we have systematically benchmarked multiple qubits across various operating frequencies, all of which consistently exhibit fidelities at the $99.99\%$ level. These comprehensive results demonstrate that high-fidelity quantum operations can be robustly achieved across different device parameters within the unified single-port architecture.

\section{Instruction-Level Waveform Synthesis for Unified Flux Control}

\begin{figure*}[bt]
  \includegraphics{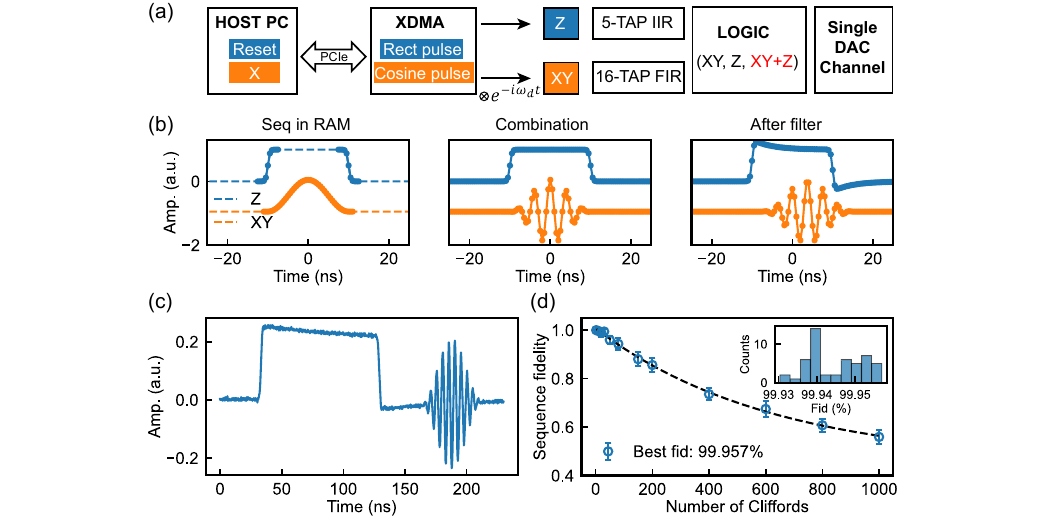}
  \caption{\textbf{Instruction-level Waveform Synthesis for Unified Flux Control.}
    (a) Control architecture. A host PC sends instruction sequences via PCIe to the AWG, where waveform primitives are stored in onboard memory. The FPGA performs real-time modulation of the $XY$ envelope at frequency $\omega_d$, applies digital filtering to the $Z$ and $XY$ paths (5-tap IIR and 16-tap FIR, respectively), and routes the resulting composite $(XY{+}Z)$ waveform to a single DAC channel.
    (b) Waveform synthesis process. Left: stored primitives, including baseband flux edges and a resonant cosine envelope. Middle: assembled and modulated waveforms prior to filtering. Right: output waveforms after digital compensation.
    (c) Measured AWG output for a composite $(XY{+}Z)$ waveform.
    (d) Randomized benchmarking of a fluxonium qubit using this control scheme. The extracted single-qubit gate fidelity is 99.957\%. Inset: distribution over repeated runs.}
  \label{fig3}
\end{figure*}

While software-level pre-distortion establishes the feasibility of unified flux control, scalability ultimately requires elevating waveform generation from stored data to dynamic synthesis. Precomputing and storing fully distorted waveforms is intrinsically non-scalable, as it ties control complexity directly to memory resources and limits runtime flexibility. A more general approach is to treat control waveforms as compositions of reusable waveform primitives, generated and conditioned during execution. 
 
Guided by this principle, we implement an FPGA-based control architecture in which the AWG operates as an instruction-driven waveform synthesizer, as shown in Fig.~\ref{fig3}(a). Instead of uploading explicit waveforms, the host provides a compact sequence of operations that reference a minimal set of reusable waveform primitives together with timing and frequency parameters~\cite{yang2022fpga}. Representative primitives, including baseband flux edges and resonant cosine envelopes, are shown in the left panel of Fig.~\ref{fig3}(b). During execution, the FPGA dynamically assembles and modulates these segments to synthesize the target control signals, as illustrated in the middle panel of Fig.~\ref{fig3}(b). In particular,the resonant envelope is modulated by a carrier $e^{-i\omega_d t}$ matched to the selected qubit, thus decoupling logical operations from their physical waveform representations. Within this framework, compensation of control-line distortions is incorporated directly into the synthesis process. The $Z$ and $XY$ paths are conditioned independently through embedded digital filters: a $5$-tap infinite impulse response (IIR) filter suppresses long-timescale distortions associated with slow flux-line dynamics, while a $16$-tap finite impulse response (FIR) filter performs dynamic pre-distortion of the microwave-modulated component~\cite{rol2020time}. Detailed characterization of these channel distortions is provided in the Appendix~\ref{distortion}. The conditioned baseband and microwave-modulated components are then summed and emitted as a single composite waveform ($XY{+}Z$) through one digital-to-analog converter (DAC). The resulting filtered waveforms are shown in the right panel of Fig.~\ref{fig3}(b). In this way, control is implemented through dynamic signal synthesis rather than stored waveform replay. 
 
Experimentally, the synthesized output waveform, including simultaneous baseband and microwave-modulated components (i.e., $XY{+}Z$), closely matches the designed structure [Fig.~\ref{fig3}(c)], with the specific FIR coefficients detailed in Appendix~\ref{fir}. To further validate the generality of this approach, we benchmark single-qubit gates on a separate fluxonium qubit operated at $202$\,MHz in a different cooldown, with measured coherence times of $T_1 \sim 40\,\mu\mathrm{s}$ and $T_2^\mathrm{R} \sim 13\,\mu\mathrm{s}$ and a gate duration of $28$\,ns. Randomized benchmarking yields a best gate fidelity of $99.957\%$ [Fig.~\ref{fig3}(d)], approaching the coherence limit. 

Notably, long gate sequences are synthesized dynamically from short primitive segments: in our implementation, sequences exceeding $50~\mu$s in duration require storing less than $100$~ns of waveform data. This abstraction leads to a substantial reduction in memory requirements. Rather than storing fully synthesized analog waveforms for every operation, the complete set of single-qubit controls can be generated from only a small library of reusable primitives. Higher-level operations, such as virtual-$Z$ phase updates, are resolved directly at the instruction level without additional waveform storage. Consequently, the program description (number of instructions and associated parameters) scales with the instruction set rather than with the stored waveform volume, enabling efficient and flexible programmability. Critically, the entire synthesized control stream is delivered through a single DAC channel, providing a direct hardware-level reduction in per-qubit control overhead.

\section{Conclusion and Outlook}

We have introduced and experimentally validated a unified flux control architecture for fluxonium qubits. The central challenge of this architecture is to reconcile three competing requirements within a single physical control channel: suppression of room-temperature wideband noise to preserve coherence, sufficient low-frequency transmission for large flux excursions and active reset, and accurate microwave delivery for high-fidelity gate operations. Our results show that these competing requirements can be resolved through the joint design of frequency-selective filtering and compensated waveform synthesis.

Specifically, frequency-selective filtering enables operation in the high-coherence regime while preserving the low-frequency control range required for initialization, and compensated pulse synthesis corrects the distortion of short microwave pulses introduced by the line response. Furthermore, to reduce electronics overhead, we implement instruction-based waveform generation on FPGA, where a compact set of reusable waveform primitives replaces stored waveform replay. Together, these elements establish a unified control framework that reduces wiring and electronics overhead while preserving coherence and enabling high-fidelity control.

This reduction in control hardware is a direct consequence of a fluxonium-specific feature: owing to its transition frequency in the hundreds-of-megahertz range, the microwave-modulated component for resonant driving and the baseband component for flux tuning can be synthesized and delivered as a single composite $(XY{+}Z)$ waveform from one AWG/DAC channel per qubit with only moderate bandwidth requirements. By contrast, the few-gigahertz transition frequency of the transmon imposes a much larger bandwidth requirement, typically forcing the use of separate channels for low-bandwidth $Z$ control and high-bandwidth microwave $XY$ drive, combined only at room temperature. This makes the transmon control stack intrinsically more hardware-intensive, whereas fluxonium allows a substantially smaller room-temperature electronics footprint per qubit.

More broadly, this work identifies a general design principle for unified flux control: passive spectral shaping and active waveform synthesis can be used as complementary tools to protect coherence while recovering controllability. This strategy is not tied to a specific filter implementation, but can be adapted to different qubit frequencies, device parameters, and line responses. A natural next step is the extension to multiplexed and multi-qubit fluxonium processors. There, the same framework could support shared-line operation through carrier multiplexing and instruction-level waveform synthesis. While larger systems will impose tighter requirements on spectral isolation, cross-talk suppression, and calibration stability, our results suggest that simplified wiring can remain compatible with high-fidelity control.

\section{Data Availability Statement}
The data that support the findings of this study are available from the corresponding author upon reasonable request.

\section*{Acknowledgments}
This research was supported by the Guangdong Provincial Quantum Science Strategic Initiative (Grant No. GDZX2407001). The authors express their gratitude to the Westlake Center for Micro/Nano Fabrication, Zhejiang QizhenTek Co., Ltd., and the Micro-Nano Fabrication and Device Center at the Songshan Lake Materials Laboratory for their essential technical support during the chip fabrication process. 

\appendix




\section{Flux line setup}
\label{mode}
We decompose the external magnetic flux into a static bias and a small time-dependent component,
\begin{equation}
\Phi(t) = \Phi_{\rm dc} + \delta\Phi(t).
\end{equation}
The static flux bias sets the operating point of the qubit and is incorporated into the Josephson term. Applying a phase-translation operator, the fluxonium Hamiltonian is then written as
\begin{equation}
H_0 = 4E_C \hat{n}^2 + \frac{1}{2}E_L \hat{\phi}^2 - E_J \cos(\hat{\phi} - \phi_{\rm dc}),
\end{equation}
where $\phi_{\rm dc} = 2\pi \Phi_{\rm dc}/\Phi_0$. The time-dependent flux $\delta\Phi(t)$ is coupled inductively through the superinductor. Introducing the reduced flux
\begin{equation}
\delta\phi(t) = 2\pi \frac{\delta\Phi(t)}{\Phi_0},
\end{equation}
the inductive term becomes
\begin{equation}
\frac{1}{2}E_L \left(\hat{\phi} - \delta\phi(t)\right)^2.
\end{equation}
Expanding to first order in $\delta\phi(t)$, we obtain
\begin{align}
\frac{1}{2}E_L \left(\hat{\phi} - \delta\phi(t)\right)^2
&= \frac{1}{2}E_L \hat{\phi}^2 
- E_L \hat{\phi}\,\delta\phi(t) + \mathcal{O}(\delta\phi^2).
\end{align}
The linear term defines the drive Hamiltonian,
\begin{equation}
\Delta H(t) = - E_L \hat{\phi}\,\delta\phi(t).
\end{equation}
The flux modulation is generated by a drive current $I_d(t)$ through a mutual inductance $M$, leading to
\begin{equation}
\delta\phi(t) = 2\pi \frac{M I_d(t)}{\Phi_0}.
\end{equation}
Substituting this into the drive Hamiltonian yields
\begin{equation}
\Delta H(t) = - E_L \cdot 2\pi \frac{M I_d(t)}{\Phi_0} \, \hat{\phi}.
\end{equation}

\subsection{Rabi Drive Strength from Room-Temperature Voltage}

To connect the drive Hamiltonian to the room-temperature control electronics, we express the drive current $I_d(t)$ in terms of the voltage generated by the arbitrary waveform generator (AWG). Let $V_{\rm awg}(t)$ be the room-temperature voltage signal. The control line includes a series of attenuators with a total voltage attenuation factor $\alpha$ (where $\alpha < 1$) and a characteristic line impedance $Z_0$ (typically $50~\Omega$). The effective current reaching the flux bias line at the base temperature is
\begin{equation}
    I_d(t) = \frac{\alpha V_{\rm awg}(t)}{Z_0}.
\end{equation}
Substituting this into the drive Hamiltonian, we obtain
\begin{equation}
    \Delta H(t) = - \left( 2\pi E_L \frac{M}{\Phi_0} \frac{\alpha}{Z_0} \right) V_{\rm awg}(t) \, \hat{\phi}.
\end{equation}
For a resonant AC drive $V_{\rm awg}(t) = V_0 \cos(\omega_d t)$, where $\omega_d$ is close to the qubit transition frequency $\omega_{ge}$ between the ground state $|0\rangle$ and the first excited state $|1\rangle$, we can apply the rotating-wave approximation (RWA). The effective drive strength, defined as the Rabi frequency $\Omega_R$, is determined by the transition matrix element of the phase operator $\langle 0 | \hat{\phi} | 1 \rangle$:
\begin{equation}
    \hbar \Omega_R = \left| 2\pi E_L \frac{M}{\Phi_0} \frac{\alpha}{Z_0} \right| V_0 \, |\langle 0 | \hat{\phi} | 1 \rangle|.
\end{equation}
This equation directly relates the room-temperature drive amplitude $V_0$ to the resulting Rabi frequency, providing a deterministic calibration for the transverse control.

\subsection{Relaxation Time Degradation Induced by Room-Temperature White Noise}

In addition to the coherent drive, the physical connection to room-temperature electronics introduces wideband noise, which can stimulate spurious transitions and degrade the qubit relaxation time $T_1$. The noise current on the flux line is characterized by its quantum noise spectral density $S_{II}(\omega)$. According to Fermi's Golden Rule, the transition rate from $|1\rangle$ to $|0\rangle$ induced by the flux line noise is given by
\begin{equation}
    \Gamma_{1\to0}^{\rm line} = \frac{1}{\hbar^2} \left( 2\pi E_L \frac{M}{\Phi_0} \right)^2 |\langle 0 | \hat{\phi} | 1 \rangle|^2 S_{II}(\omega_{ge}).
\end{equation}
Assuming the dominant noise source originates from the room-temperature electronics (e.g., AWG voltage white noise or thermal noise at $300$~K) that propagates down the control line, the effective voltage noise spectral density at the AWG output is denoted as $S_{VV}^{\rm awg}(\omega)$. After passing through the attenuation chain, the current noise spectral density at the qubit plane becomes
\begin{equation}
    S_{II}(\omega_{ge}) = \frac{\alpha^2}{Z_0^2} S_{VV}^{\rm awg}(\omega_{ge}).
\end{equation}
Substituting the current noise spectral density into the transition rate, the noise-induced relaxation rate $\Gamma_{1\to0}^{\rm line} = 1/T_1^{\rm line}$ is
\begin{equation}
    \frac{1}{T_1^{\rm line}} = \frac{1}{\hbar^2} \left( 2\pi E_L \frac{M}{\Phi_0} \frac{\alpha}{Z_0} \right)^2 |\langle 0 | \hat{\phi} | 1 \rangle|^2 S_{VV}^{\rm awg}(\omega_{ge}).
\end{equation}
This result highlights a fundamental physical trade-off in the unified control architecture: while increasing the effective coupling (by increasing $M$ or decreasing the attenuation $\alpha$) enhances the Rabi frequency $\Omega_R$ for faster gate operations, it simultaneously quadratically increases the qubit's susceptibility to room-temperature wideband noise, thereby suppressing the $T_1$ limit. This strictly necessitates the deployment of frequency-selective filtering to suppress $S_{VV}^{\rm awg}(\omega_{ge})$ while maintaining high transparency at low frequencies. To evaluate the specific impact of room-temperature electronics, we model the wideband fluctuations as frequency-independent white noise. The effective voltage noise spectral density $S_{VV}^{\rm awg}(\omega)$ at the input of the control line is typically dominated by two sources: the Johnson--Nyquist thermal noise at room temperature ($T_{\rm rt} = 300$~K) and the intrinsic noise floor of the AWG. The double-sided thermal noise spectral density is given by
\begin{equation}
    S_{VV}^{\rm th}(\omega) = 2 k_B T_{\rm rt} Z_0,
\end{equation}
where $k_B$ is the Boltzmann constant. In practical experimental setups, the active noise floor of the AWG is often significantly higher than the $300$~K thermal background. If the AWG exhibits a constant white noise power spectral density $P_{\rm noise}$ (in W/Hz), the corresponding double-sided voltage noise spectral density becomes
\begin{equation}
    S_{VV}^{\rm awg}(\omega) = \frac{1}{2} P_{\rm noise} Z_0.
\end{equation}
Substituting this effective $S_{VV}^{\rm awg}(\omega_{ge})$ back into the transition rate equation allows us to quantitatively predict the Purcell-like $T_1$ degradation induced by the specific unified-drive electronics.  It is important to note that in our main text analysis, we exclusively consider the input noise from the AWG, as the technical noise floor of the room-temperature electronics overwhelmingly dominates the standard Johnson-Nyquist thermal noise.

\section{Experimental Setup}
\label{Setup}

\begin{figure}
  \includegraphics[width = 0.48\textwidth]{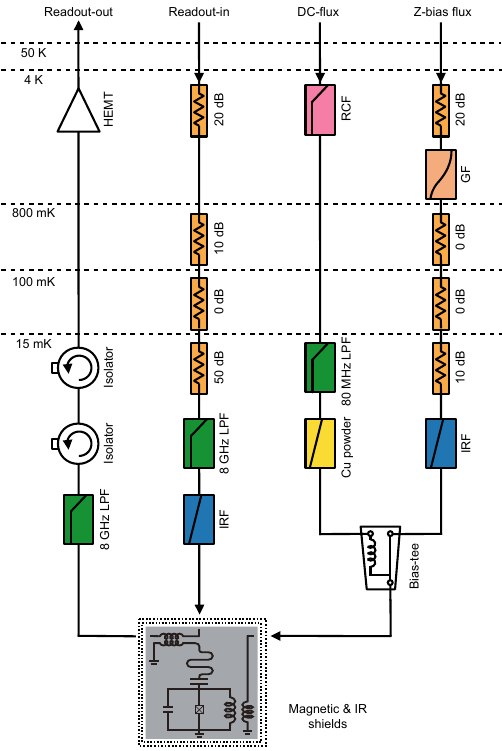} \caption{\label{experiment_setup}\textbf{Cryogenic Setup.} Schematic detailing the low-temperature wiring inside the dilution refrigerator for unified control.}
\end{figure}

The cryogenic wiring is detailed in Fig.~\ref{experiment_setup}. The device is mounted at the $15$\,mK base stage of a dilution refrigerator, enclosed within multiple layers of magnetic and infrared (IR) shielding. The setup consists of standard readout input/output lines and two flux lines: a unified dynamic control line and a static DC-flux line. The readout input line is thermalized with $80$\,dB of distributed attenuation, while the output line features cascaded isolators and a high-electron-mobility transistor (HEMT) amplifier at $4$\,K to optimize the signal-to-noise ratio and protect the qubit from back-action.

The unified dynamic control line mediates both the transverse microwave drive ($XY$ control) and the dynamic longitudinal tuning ($Z$ control). To balance drive strength and coherence, this line is equipped with $30$\,dB of cryogenic attenuation. Crucially, a Gaussian low-pass filter with a $92$\,MHz 3-dB cutoff is installed at the $4$\,K stage to strictly suppress wideband noise near the qubit transition frequency.The static DC-flux line is heavily low-pass filtered to set the baseline operating point (e.g., the half-flux sweet spot). The dynamic and static flux signals are subsequently combined via a cryogenic bias-tee at the base stage before coupling to the qubit. Notably, this dedicated DC-flux line is used here for experimental convenience; in future scaled-up architectures, it can be entirely replaced by a global magnetic field bias, leaving the unified dynamic line as the sole local routing overhead per qubit.

\section{Coherence Analysis}
\label{coherence}

To characterize the decoherence mechanisms of the unified single-port architecture, we measure the energy relaxation time $T_1$ and the pure dephasing time $T_\phi$ of the fluxonium qubit as a function of the external magnetic flux $\Phi_{\rm ext}$. During these measurements, the qubit is idled at various flux biases using the dynamic $Z$-control channel, while all microwave excitation and readout operations are performed at the half-flux sweet spot ($\Phi_{\rm ext} = 0.5 \Phi_0$).

For each flux bias $\Phi_{\rm ext}$, the measured excited-state population $P_e(t)$ is first fitted using a double-exponential relaxation model. This model accounts for deviations from a simple single-exponential decay, which can arise from quasiparticle-related relaxation dynamics~\cite{pop2014coherent,Serniak_2018}. The fitted population decay is described by
\begin{equation}
    P_e(t) = A \exp(-t/T_{\rm exp}) \exp\left\{ n_{\rm qp} \left[ \exp(-t/T_{\rm qp}) - 1 \right] \right\} + B,
\end{equation}
where $T_{\rm exp}$ and $T_{\rm qp}$ are characteristic relaxation times, $n_{\rm qp}$ is the quasiparticle density parameter, $A$ is the initial amplitude, and $B$ accounts for the measurement baseline. Because the decay is non-exponential, we do not report a single fitted rate parameter. Instead, we define an effective relaxation time $T_1$ as the time at which the fitted relaxation curve reaches the $1/e$ population level relative to its initial amplitude. This approach provides a consistent metric for energy relaxation across the entire flux tuning range.

To extract the pure dephasing time $T_\phi$, we perform Ramsey, Hahn echo, and Carr-Purcell-Meiboom-Gill (CPMG) measurements. The envelope of the state projection signal $P_{env}(t)$ in these transverse relaxation measurements is fitted to a mixed exponential-plus-Gaussian decay function~\cite{wang2024efficient,Ithier_2005,Bylander_2011}:
\begin{equation}
    P_{env}(t) = C \exp\left( -\frac{t}{2T_1^{\rm DE}} \right) \exp\left[ -\frac{t}{T_{\phi,\rm exp}} - \left(\frac{t}{T_{\phi,\rm G}}\right)^2 \right] + D.
\end{equation}
Here, the term $\exp[-t/(2T_1^{\rm DE})]$ represents the fixed relaxation contribution, where $T_1^{\rm DE}$ denotes the relaxation time scale derived directly from the double-exponential (DE) fit to the $T_1$ data, rather than the effective $1/e$ time. The pure dephasing is then characterized by both an exponential time constant $T_{\phi,\rm exp}$ and a Gaussian time constant $T_{\phi,\rm G}$. Away from the sweet spot, low-frequency $1/f$ flux noise becomes the dominant dephasing channel, which manifests as a Gaussian decay profile. Therefore, in our analysis, we extract the Gaussian component $T_\phi \equiv T_{\phi,\rm G}$ as the relevant pure-dephasing time.

\begin{figure}
  \includegraphics[width = 0.48\textwidth]{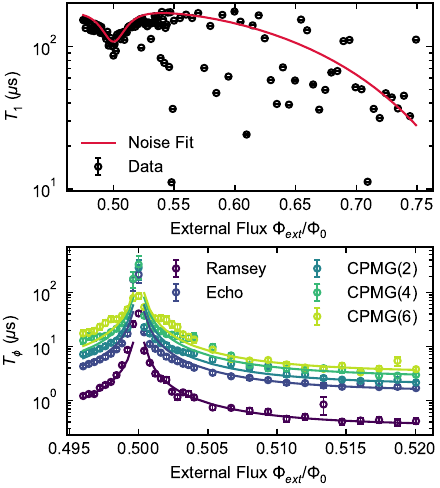} \caption{\label{coherence_vs_flux}\textbf{Coherence vs. Flux.} 
  (a) Energy relaxation time $T_1$ as a function of the external flux. 
  (b) Pure dephasing time $T_\phi$ extracted from Ramsey, Hahn echo, and CPMG sequences with varying numbers of refocusing $\pi$ pulses as a function of the external flux.}
\end{figure}

As shown in Fig.~\ref{coherence_vs_flux}(a), the effective energy relaxation time $T_1$ is generally maintained between $100$ and $200~\mu$s near the half-flux sweet spot. When the flux is tuned away from this optimal point, $T_1$ exhibits noticeable fluctuations and several sharp, precipitous drops (e.g., near $\Phi_{\rm ext}/\Phi_0 \approx 0.55$). These narrow-band suppressions of the relaxation time are characteristic signatures of resonant interactions with randomly distributed two-level system (TLS) defects in the device environment~\cite{nguyen2019high,bao2022fluxonium,Martinis_2005,Muller_2019}. 

The flux dependence of the extracted pure dephasing time $T_\phi$ is presented in Fig.~\ref{coherence_vs_flux}(b). At the half-flux sweet spot ($\Phi_{\rm ext} = 0.5 \Phi_0$), the qubit transition frequency is first-order insensitive to flux variations ($d\omega_{ge}/d\Phi_{\rm ext} = 0$), leading to a strongly peaked $T_\phi$ that exceeds $100~\mu$s. As the flux bias moves away from the sweet spot, the qubit frequency becomes highly sensitive to local magnetic flux fluctuations, causing $T_\phi$ to drop rapidly to the single-microsecond regime. This steep decline is primarily driven by the aforementioned low-frequency $1/f$ flux noise~\cite{Yoshihara_2006,Paladino_2014}. By employing dynamical decoupling techniques, the low-frequency phase accumulation can be effectively mitigated. As shown in Fig.~\ref{coherence_vs_flux}(b), increasing the number of refocusing $\pi$ pulses in the CPMG($N$) sequences progressively filters out the low-frequency noise components, significantly extending $T_\phi$ away from the sweet spot. 

To further identify the physical origins of decoherence, we perform independent fits to the flux-dependent $T_1$ and $T_\phi$ data using a comprehensive noise model. The resulting fit curves are shown as solid lines in Fig.~\ref{coherence_vs_flux}. We first fit the pure dephasing data to a model dominated by low-frequency $1/f$ flux noise and a frequency-independent white noise floor, yielding a $1/f$ flux noise amplitude of $A_\Phi \approx \text{8.11}\pm\text{0.09}~\mu\Phi_0$ and a white flux noise amplitude of $A_w \approx \text{4.18}\pm\text{0.27}~n\Phi_0/\sqrt{\text{Hz}}$. Using the extracted $1/f$ flux noise intensity as a fixed parameter, we then fit the relaxation data to a model incorporating dielectric loss from TLS, $1/f$ flux noise, and an additional free static noise term. This procedure yields an effective dielectric loss tangent of $\tan \delta_C \approx \text{13.7} \times 10^{-6}$. These quantitative results confirm that the qubit coherence is governed by the interplay between TLS-induced dielectric loss and magnetic flux fluctuations.

\section{Summary of All Tested Devices}
\label{all_devices}

The same approach has also been tested on multiple additional devices, as summarized in Table~\ref{tab:all_results}. Device A corresponds to the qubit discussed in the main text. Devices B, C, D, and E all support high-fidelity control, with extracted single-qubit gate fidelities above 99.97\%. Devices F and G were not fully benchmarked by randomized benchmarking, but both support standard gate operations and exhibit coherence times with $T_1 > 100~\mu$s. Device H, with a substantially higher qubit frequency, can still be operated under the same control architecture, but the 92~MHz low-pass filter requires significantly longer pulses, with gate durations extended to about 200~ns. Taken together, these results indicate that a single 92~MHz low-pass filter is sufficient to support high-fidelity control for fluxonium qubits with transition frequencies below approximately 300~MHz.

More broadly, these results clarify both the applicability and the limitation of the present design. For qubits in the sub-300~MHz range, strong low-pass filtering can simultaneously preserve long coherence and support accurate compensated control. At higher qubit frequencies, the same filtering strategy remains conceptually valid, but maintaining short and high-fidelity gates requires correspondingly larger filter bandwidth. This does not alter the main conclusion of this work. Rather, it reinforces the central design principle: unified single-port control is enabled not by a specific cutoff frequency, but by the joint use of frequency-selective filtering and waveform compensation. 

\begin{table}[h]
\centering
\caption{Results of all tested devices.}
\label{tab:all_results}
\setlength{\tabcolsep}{4pt}
\resizebox{0.48\textwidth}{!}{%
\begin{tabular}{c|cccccc}
\toprule
Device & $f_{q}$(MHz) & Fidelity (\%) & $T_g$ (ns)& $T_1$($\mu$s) & $T_2^R$($\mu$s) & $T_2^{\mathrm{echo}}$($\mu$s) \\
\midrule
A & 208 & 99.990 & 20 & 110 & 128 &133 \\
B & 205 & 99.974 & 20 & 65 & 21 &34 \\
C & 232 & 99.986 & 20 & 95 & 82 &118 \\
D & 285 & 99.970 & 40 & 53 & 30 &34 \\
E & 267 & 99.974 & 30 & 52 & 23 &46 \\
F & 233 & N/A & 20 & 140 & 28 &28 \\
G & 164 & N/A & 20 & 151 & 30 &79 \\
H & 378 & N/A & 200 & 214 & 68 &131 \\
\bottomrule
\end{tabular}%
}
\end{table}

\section{Distortion characterization}
\label{distortion}

\begin{figure}
  \includegraphics{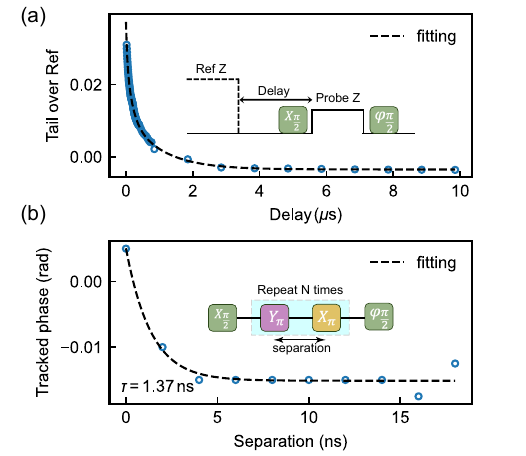}     \caption{\label{fig:zpulse_distortion}\textbf{Distortion Characterization in the Unified Single-port Architecture.}
    (a) Long-tail response of the $Z$ control channel, measured by converting the phase accumulated during a delayed probe window into the equivalent normalized residual $Z$-pulse tail. The dashed line is an exponential fit used for IIR-based compensation.
    (b) Assessment of residual microwave intra-quadrature distortion using alternating $Y_{\pi}$ and $X_{\pi}$ pulses. The dashed line is a single-exponential fit with a characteristic decay time of $1.37$~ns, indicating that the residual tail is sufficiently short that no further correction is applied.}
\end{figure}

To assess the dominant waveform imperfections in the unified single-port architecture, we separately characterize the settling dynamics of the low-frequency $Z$ channel and the residual temporal distortion in the microwave control channel. The corresponding measurements are summarized in Fig.~\ref{fig:zpulse_distortion}. For the $Z$ channel, the dominant imperfection is a long-tail response following the falling edge of a nominally rectangular flux pulse. Such settling behavior may arise from impedance mismatch, surface-related relaxation, or effective LR-type filtering in the flux-control path, and it can introduce unwanted phase accumulation in subsequent qubit operations. To quantify this effect, we use the qubit as an in-situ detector. As shown in Fig.~\ref{fig:zpulse_distortion}(a), a long reference $Z$ pulse is first applied to generate a well-defined falling edge. After a variable delay, a short probe pulse shifts the qubit to a flux-sensitive point, where the residual tail accumulated during the probe window is converted into a measurable phase shift through a Ramsey-type sequence~\cite{bao2022fluxonium}. In the present analysis, rather than deconvolving the full flux-distortion waveform from the measured phase, we directly convert the phase into the equivalent residual $Z$-pulse tail normalized to the reference-pulse amplitude. The vertical axis in Fig.~\ref{fig:zpulse_distortion}(a) therefore directly represents the relative settling tail, denoted as ``Tail over Ref''.

The extracted $Z$-pulse tail is well described by a multi-exponential relaxation model. From the fit, we obtain three settling times of approximately $34$~ns, $170$~ns, and $996$~ns, with corresponding normalized amplitudes of $-1.74\%$, $-1.89\%$, and $-1.58\%$, respectively. These results indicate that the $Z$-line distortion contains both intermediate and long-timescale components, consistent with slow settling processes in the flux-control path. The extracted parameters provide a compact description of the low-frequency response and directly motivate the use of IIR-based compensation for the $Z$ channel.

Microwave pulse distortions can in general be divided into intra-quadrature distortion and cross-quadrature distortion. For an ideal resonant drive, the effective Hamiltonian in the rotating frame is
\begin{equation}
H_{\rm ideal}(t)=\frac{\Omega(t)}{2}\sigma_x,
\end{equation}
where $\Omega(t)$ is the target pulse envelope. Intra-quadrature distortion corresponds to residual errors that remain in the same quadrature channel, leading to
\begin{equation}
H_{\rm intra}(t)=\frac{\Omega(t)+\delta\Omega_x(t)}{2}\sigma_x,
\end{equation}
which mainly gives rise to coherent over- or under-rotation without changing the nominal rotation axis. By contrast, cross-quadrature distortion introduces an orthogonal component,
\begin{equation}
H_{\rm cross}(t)=\frac{\Omega(t)}{2}\sigma_x+\frac{\delta\Omega_y(t)}{2}\sigma_y,
\end{equation}
thereby tilting the effective rotation axis in the rotating frame.

In our experiment, qubit control is implemented using direct microwave flux drive. The flux-drive operator predominantly couples the $\ket{g}\leftrightarrow\ket{e}$ transition, while its coupling to higher transitions is strongly suppressed. As a result, temporal waveform imperfections are expected to remain predominantly within the effective control quadrature. Nevertheless, a residual tail left by a preceding pulse can still appear as a small orthogonal correction during the following operation and therefore can be detected through its induced phase shift.

To characterize this effect, we employ alternating $Y_\pi$ and $X_\pi$ pulses, as shown in Fig.~\ref{fig:zpulse_distortion}(b). Physically, a distorted $Y_\pi$ pulse leaves a weak temporal tail that persists into the subsequent $X_\pi$ pulse. The effective Hamiltonian during the $X_\pi$ operation can therefore be written as
\begin{equation}
H_{\rm eff}
=
\frac{\Omega_x(t)}{2}\sigma_x
+
\frac{\delta\Omega_y(t-t_d)}{2}\sigma_y,
\end{equation}
where $t_d$ is the delay between neighboring pulses. The residual tail tilts the effective rotation axis and produces a small coherent phase shift,
\begin{equation}
\phi_s \approx \frac{\delta\Omega_y}{\Omega_x}.
\end{equation}
This phase accumulation can be measured in a Ramsey-type interference experiment by scanning both the pulse separation and the phase of the final analysis pulse.

As shown in Fig.~\ref{fig:zpulse_distortion}(b), the measured phase is well described by a single-exponential decay with a characteristic time of only $1.37$~ns. This timescale is much shorter than the gate duration relevant for the present experiment, indicating that the residual microwave tail is already weak and rapidly decaying. Accordingly, we do not apply any further correction based on this measurement. Instead, this experiment serves only as an assessment of the residual microwave distortion level, rather than as an input for determining the FIR coefficients. Taken together with the $Z$-channel measurement, these results indicate that the dominant correctable settling error in our setup arises from the low-frequency flux path, while the residual microwave tail is sufficiently short that no additional correction is required.

\section{FIR coefficient synthesis for microwave compensation}
\label{fir}

\begin{figure}
  \includegraphics{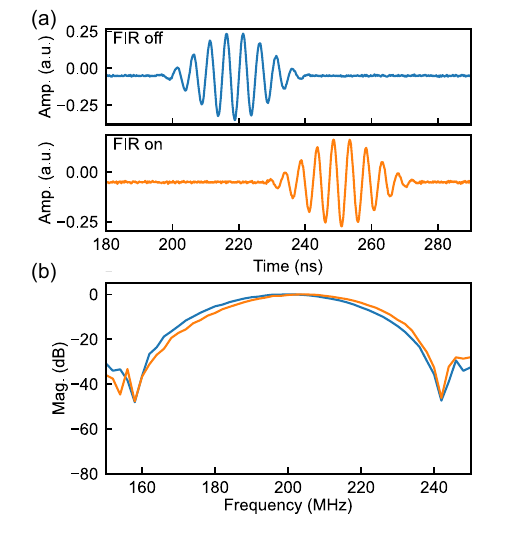}     \caption{\label{fig:fir_on_off}\textbf{FIR Coefficient Synthesis and Measured Waveform Reshaping.}
    (a) Measured output waveforms with the FIR block disabled (top) and enabled (bottom). The 16-tap FIR coefficients are generated from the target bounded inverse transfer function by frequency-domain sampling, linear-phase filter synthesis, and 16-bit quantization for FPGA implementation. With FIR enabled, the waveform is visibly reshaped and shifted in time, reflecting the causal group delay and spectral reweighting introduced by the digital filter.
    (b) Measured spectra of the corresponding waveforms with FIR off (blue) and FIR on (orange). The FIR-on waveform exhibits reduced low-frequency spectral weight and enhanced higher-frequency components around the qubit control band, consistent with the intended pre-compensation of the downstream low-pass response.}
\end{figure}

The FPGA FIR coefficients are synthesized directly from the compensated transfer function $H_{\rm inv}(f)$ obtained in the main text. First, the target gain profile is sampled on a dense frequency grid from dc to the Nyquist frequency,
\begin{equation}
G(f)=|H_{\rm inv}(f)|.
\end{equation}
The sampled frequency response is then used as the design target for a 16-tap linear-phase FIR filter using frequency-domain synthesis. Since an even number of symmetric taps is adopted, the filter belongs to the Type-II FIR class, whose response is constrained to vanish at the Nyquist frequency. Accordingly, the target gain at the Nyquist frequency is explicitly set to zero before the synthesis procedure.

After obtaining the floating-point FIR coefficients $h_i$, the coefficients are normalized to the maximum absolute value and quantized into signed 16-bit integers for FPGA implementation,
\begin{equation}
h_i^{(\mathrm{int16})}
=
\mathrm{round}
\left(
\frac{h_i}{\max|h_i|}
\times 32767
\right).
\end{equation}
The final FIR coefficients used in the experiment are $h_{\rm FIR} =
[-300,\ 145,\ 977,\ 3351,\ -682,\ -9355,\ -25925,\ 32767,\ \\ 32766,\ -25925,\ -9355,\ -682,\ 3351,\ 977,\ 145,\ -300]$. 

The effect of this digital filter is shown in Fig.~\ref{fig:fir_on_off}. In the time domain, enabling the FIR block visibly reshapes the output waveform and introduces a temporal shift relative to the FIR-off case. This shift is the expected consequence of causal digital filtering and reflects the finite group delay of the FIR process. At the same time, the envelope is redistributed in a way consistent with pre-compensation. In the frequency domain, the FIR-on waveform exhibits suppressed low-frequency spectral weight and relatively enhanced higher-frequency components around the qubit control band. This spectral redistribution is precisely the intended function of the digital compensation: rather than simply amplifying the entire waveform, the FIR selectively suppresses the slowly varying part of the signal while boosting the faster components needed to counteract the downstream low-pass response. In this way, the FIR block provides a hardware-compatible implementation of the compensated microwave waveform synthesis discussed in the main text.



\bibliographystyle{apsrev4-2}
\bibliography{references_with_all_links}

@article{van2024advanced,
  author = {Van Damme, J and Massar, S and Acharya, R and Ivanov, Ts and Perez Lozano, D and Canvel, Y and Demarets, M and Vangoidsenhoven, D and Hermans, Y and Lai, JG and others},
  title = {Advanced {CMOS} manufacturing of superconducting qubits on 300 mm wafers},
  journal = {Nature},
  volume = {634},
  number = {8032},
  pages = {74--79},
  year = {2024},
  doi = {10.1038/s41586-024-07941-9},
  url = {https://doi.org/10.1038/s41586-024-07941-9},
}

@article{arute2019quantum,
  author = {Arute, Frank and Arya, Kunal and Babbush, Ryan and Bacon, Dave and Bardin, Joseph C and Barends, Rami and Biswas, Rupak and Boixo, Sergio and Brandao, Fernando GSL and Buell, David A. and others},
  title = {Quantum Supremacy Using a Programmable Superconducting Processor},
  journal = {Nature},
  volume = {574},
  number = {7779},
  pages = {505--510},
  year = {2019},
  doi = {10.1038/s41586-019-1666-5},
  url = {https://doi.org/10.1038/s41586-019-1666-5},
}

@article{Wu2021,
  author = {Wu, Yulin and Bao, Wan-Su and Cao, Sirui and Chen, Fusheng and Chen, Ming-Cheng and Chen, Xiawei and Chung, Tung-Hsun and Deng, Hui and Du, Yajie and Fan, Daojin and Gong, Ming and Guo, Cheng and Guo, Chu and Guo, Shaojun and Han, Lianchen and Hong, Linyin and Huang, He-Liang and Huo, Yong-Heng and Li, Liping and Li, Na and others},
  title = {Strong Quantum Computational Advantage Using a Superconducting Quantum Processor},
  journal = {Phys. Rev. Lett.},
  volume = {127},
  number = {18},
  pages = {180501},
  year = {2021},
  doi = {10.1103/PhysRevLett.127.180501},
  url = {https://link.aps.org/doi/10.1103/PhysRevLett.127.180501},
}

@article{kim2023evidence,
  author = {Kim, Youngseok and Eddins, Andrew and Anand, Sajant and Wei, Ken Xuan and Van Den Berg, Ewout and Rosenblatt, Sami and Nayfeh, Hasan and Wu, Yantao and Zaletel, Michael and Temme, Kristan and others},
  title = {Evidence for the Utility of Quantum Computing Before Fault Tolerance},
  journal = {Nature},
  volume = {618},
  number = {7965},
  pages = {500--505},
  year = {2023},
  doi = {10.1038/s41586-023-06096-3},
  url = {https://doi.org/10.1038/s41586-023-06096-3},
}

@article{Jin2025,
  author = {Jin, Feitong and Jiang, Si and Zhu, Xuhao and Bao, Zehang and Shen, Fanhao and Wang, Ke and Zhu, Zitian and Xu, Shibo and Song, Zixuan and Chen, Jiachen and Tan, Ziqi and Wu, Yaozu and Zhang, Chuanyu and Gao, Yu and Wang, Ning and Zou, Yiren and Zhang, Aosai and Li, Tingting and Zhong, Jiarun and Cui, Zhengyi and others},
  title = {Topological Prethermal Strong Zero Modes on Superconducting Processors},
  journal = {Nature},
  volume = {645},
  number = {8081},
  pages = {626--632},
  year = {2025},
  doi = {10.1038/s41586-025-09476-z},
  url = {https://doi.org/10.1038/s41586-025-09476-z},
}

@article{Koch2007,
  author = {Koch, Jens and Yu, Terri M and Gambetta, Jay and Houck, Andrew A and Schuster, David I and Majer, Johannes and Blais, Alexandre and Devoret, Michel H and Girvin, Steven M and Schoelkopf, Robert J},
  title = {Charge-insensitive Qubit Design Derived from the Cooper Pair Box},
  journal = {Phys. Rev. A},
  volume = {76},
  number = {4},
  pages = {042319},
  year = {2007},
  doi = {10.1103/PhysRevA.76.042319},
  url = {https://doi.org/10.1103/PhysRevA.76.042319},
}

@article{pop2014coherent,
  author = {Pop, Ioan M and Geerlings, Kurtis and Catelani, Gianluigi and Schoelkopf, Robert J and Glazman, Leonid I and Devoret, Michel H},
  title = {Coherent suppression of electromagnetic dissipation due to superconducting quasiparticles},
  journal = {Nature},
  volume = {508},
  number = {7496},
  pages = {369--372},
  year = {2014},
  doi = {10.1038/nature13017},
  url = {https://doi.org/10.1038/nature13017},
}

@article{manucharyan2009fluxonium,
  author = {Manucharyan, Vladimir E. and Koch, Jens and Glazman, Leonid I. and Devoret, Michel H.},
  title = {{Fluxonium}: Single {{Cooper}}-Pair Circuit Free of Charge Offsets},
  journal = {Science},
  volume = {326},
  pages = {113--116},
  year = {2009},
  doi = {10.1126/science.1175552},
  url = {https://doi.org/10.1126/science.1175552},
}

@article{nguyen2019high,
  author = {Nguyen, Long B. and Lin, Yen-Hsiang and Somoroff, Aaron and Mencia, Raymond and Grabon, Nicholas and Manucharyan, Vladimir E.},
  title = {High-Coherence {Fluxonium} Qubit},
  journal = {Phys. Rev. X},
  volume = {9},
  number = {4},
  pages = {041041},
  year = {2019},
  doi = {10.1103/PhysRevX.9.041041},
  url = {https://link.aps.org/doi/10.1103/PhysRevX.9.041041},
}

@article{Somoroff2023,
  author = {Somoroff, Aaron and Ficheux, Quentin and Mencia, Raymond A. and Xiong, Haonan and Kuzmin, Roman and Manucharyan, Vladimir E.},
  title = {Millisecond Coherence in a Superconducting Qubit},
  journal = {Phys. Rev. Lett.},
  volume = {130},
  number = {26},
  pages = {267001},
  year = {2023},
  doi = {10.1103/PhysRevLett.130.267001},
  url = {https://link.aps.org/doi/10.1103/PhysRevLett.130.267001},
}

@article{Wang2025,
  author = {Wang, Fei and Lu, Kannan and Zhan, Huijuan and Ma, Lu and Wu, Feng and Sun, Hantao and Deng, Hao and Bai, Yang and Bao, Feng and Chang, Xu and Gao, Ran and Gao, Xun and Gong, Guicheng and Hu, Lijuan and Hu, Ruizi and Ji, Honghong and Ma, Xizheng and Mao, Liyong and Song, Zhijun and Tang, Chengchun and others},
  title = {High-Coherence {Fluxonium} Qubits Manufactured with a Wafer-Scale-Uniformity Process},
  journal = {Phys. Rev. Appl.},
  volume = {23},
  number = {4},
  pages = {044064},
  year = {2025},
  doi = {10.1103/PhysRevApplied.23.044064},
  url = {https://link.aps.org/doi/10.1103/PhysRevApplied.23.044064},
}

@article{Zhang2021Universal,
  author = {Zhang, Helin and Chakram, Srivatsan and Roy, Tanay and Earnest, Nathan and Lu, Yao and Huang, Ziwen and Weiss, D. K. and Koch, Jens and Schuster, David I.},
  title = {Universal Fast-Flux Control of a Coherent, Low-Frequency Qubit},
  journal = {Phys. Rev. X},
  volume = {11},
  number = {1},
  pages = {011010},
  year = {2021},
  doi = {10.1103/PhysRevX.11.011010},
  url = {https://doi.org/10.1103/PhysRevX.11.011010},
}

@article{Ficheux2021,
  author = {Ficheux, Quentin and Nguyen, Long B. and Somoroff, Aaron and Xiong, Haonan and Nesterov, Konstantin N. and Vavilov, Maxim G. and Manucharyan, Vladimir E.},
  title = {Fast Logic with Slow Qubits: Microwave-Activated Controlled-Z Gate on Low-Frequency {Fluxoniums}},
  journal = {Phys. Rev. X},
  volume = {11},
  number = {2},
  pages = {021026},
  year = {2021},
  doi = {10.1103/PhysRevX.11.021026},
  url = {https://link.aps.org/doi/10.1103/PhysRevX.11.021026},
}

@article{bao2022fluxonium,
  author = {Bao, Feng and Deng, Hao and Ding, Dawei and Gao, Ran and Gao, Xun and Huang, Cupjin and Jiang, Xun and Ku, Hsiang-Sheng and Li, Zhisheng and Ma, Xizheng and Ni, Xiaotong and Qin, Jin and Song, Zhijun and Sun, Hantao and Tang, Chengchun and Wang, Tenghui and Wu, Feng and Xia, Tian and Yu, Wenlong and Zhang, Fang and others},
  title = {{Fluxonium}: An Alternative Qubit Platform for High-Fidelity Operations},
  journal = {Phys. Rev. Lett.},
  volume = {129},
  number = {1},
  pages = {010502},
  year = {2022},
  doi = {10.1103/PhysRevLett.129.010502},
  url = {https://link.aps.org/doi/10.1103/PhysRevLett.129.010502},
}

@article{huang2023quantum,
  author = {Huang, Cupjin and Wang, Tenghui and Wu, Feng and Ding, Dawei and Ye, Qi and Kong, Linghang and Zhang, Fang and Ni, Xiaotong and Song, Zhijun and Shi, Yaoyun and Zhao, Hui-Hai and Deng, Chunqing and Chen, Jianxin},
  title = {Quantum Instruction Set Design for Performance},
  journal = {Phys. Rev. Lett.},
  volume = {130},
  number = {7},
  pages = {070601},
  year = {2023},
  doi = {10.1103/PhysRevLett.130.070601},
  url = {https://link.aps.org/doi/10.1103/PhysRevLett.130.070601},
}

@article{ma2023native,
  author = {Ma, Xizheng and Zhang, Gengyan and Wu, Feng and Bao, Feng and Chang, Xu and Chen, Jianjun and Deng, Hao and Gao, Ran and Gao, Xun and Hu, Lijuan and Ji, Honghong and Ku, Hsiang-Sheng and Lu, Kannan and Ma, Lu and Mao, Liyong and Song, Zhijun and Sun, Hantao and Tang, Chengchun and Wang, Fei and Wang, Hongcheng and others},
  title = {Native Approach to Controlled-{$Z$} Gates in Inductively Coupled {Fluxonium} Qubits},
  journal = {Phys. Rev. Lett.},
  volume = {132},
  number = {6},
  pages = {060602},
  year = {2024},
  doi = {10.1103/PhysRevLett.132.060602},
  url = {https://link.aps.org/doi/10.1103/PhysRevLett.132.060602},
}

@article{ding2023high,
  author = {Ding, Leon and Hays, Max and Sung, Youngkyu and Kannan, Bharath and An, Junyoung and Di Paolo, Agustin and Karamlou, Amir H. and Hazard, Thomas M. and Azar, Kate and Kim, David K. and Niedzielski, Bethany M. and Melville, Alexander and Schwartz, Mollie E. and Yoder, Jonilyn L. and Orlando, Terry P. and Gustavsson, Simon and Grover, Jeffrey A. and Serniak, Kyle and Oliver, William D.},
  title = {High-Fidelity, Frequency-Flexible Two-Qubit {Fluxonium} Gates with a Transmon Coupler},
  journal = {Phys. Rev. X},
  volume = {13},
  number = {3},
  pages = {031035},
  year = {2023},
  doi = {10.1103/PhysRevX.13.031035},
  url = {https://link.aps.org/doi/10.1103/PhysRevX.13.031035},
}

@article{Reed2010fast,
  author = {Reed, M. D. and Johnson, B. R. and Houck, A. A. and DiCarlo, L. and Chow, J. M. and Schuster, D. I. and Frunzio, L. and Schoelkopf, R. J.},
  title = {{Fast Reset and Suppressing Spontaneous Emission of a Superconducting Qubit}},
  journal = {Appl. Phys. Lett.},
  volume = {96},
  number = {20},
  pages = {203110},
  year = {2010},
  doi = {10.1063/1.3435463},
  url = {https://doi.org/10.1063/1.3435463},
}

@article{mcewen2021removing,
  author = {McEwen, Matt and Kafri, Dvir and Chen, Z and Atalaya, Juan and Satzinger, KJ and Quintana, Chris and Klimov, Paul Victor and Sank, Daniel and Gidney, C and Fowler, AG and others},
  title = {Removing Leakage-Induced Correlated Errors in Superconducting Quantum Error Correction},
  journal = {Nat. Commun.},
  volume = {12},
  number = {1},
  pages = {1761},
  year = {2021},
  doi = {10.1038/s41467-021-21982-y},
  url = {https://doi.org/10.1038/s41467-021-21982-y},
}

@article{magesan2012efficient,
  author = {Magesan, Easwar and Gambetta, Jay M and Johnson, Blake R and Ryan, Colm A and Chow, Jerry M and Merkel, Seth T and Da Silva, Marcus P and Keefe, George A and Rothwell, Mary B and Ohki, Thomas A and others},
  title = {Efficient Measurement of Quantum Gate Error by Interleaved Randomized Benchmarking},
  journal = {Phys. Rev. Lett.},
  volume = {109},
  number = {8},
  pages = {080505},
  year = {2012},
  doi = {10.1103/PhysRevLett.109.080505},
  url = {https://doi.org/10.1103/PhysRevLett.109.080505},
}

@article{Negirneac2021,
  author = {Neg\^{\i}rneac, V. and Ali, H. and Muthusubramanian, N. and Battistel, F. and Sagastizabal, R. and Moreira, M. S. and Marques, J. F. and Vlothuizen, W. J. and Beekman, M. and Zachariadis, C. and Haider, N. and Bruno, A. and DiCarlo, L.},
  title = {High-Fidelity Controlled-$Z$ Gate with Maximal Intermediate Leakage Operating at the Speed Limit in a Superconducting Quantum Processor},
  journal = {Phys. Rev. Lett.},
  volume = {126},
  number = {22},
  pages = {220502},
  year = {2021},
  doi = {10.1103/PhysRevLett.126.220502},
  url = {https://link.aps.org/doi/10.1103/PhysRevLett.126.220502},
}

@article{ganjam2024surpassing,
  author = {Ganjam, Suhas and Wang, Yanhao and Lu, Yao and Banerjee, Archan and Lei, Chan U and Krayzman, Lev and Kisslinger, Kim and Zhou, Chenyu and Li, Ruoshui and Jia, Yichen and others},
  title = {Surpassing millisecond coherence in on chip superconducting quantum memories by optimizing materials and circuit design},
  journal = {Nat. Commun.},
  volume = {15},
  number = {1},
  pages = {3687},
  year = {2024},
  doi = {10.1038/s41467-024-47857-6},
  url = {https://doi.org/10.1038/s41467-024-47857-6},
}

@article{place2021new,
  author = {Place, Alexander PM and Rodgers, Lila VH and Mundada, Pranav and Smitham, Basil M and Fitzpatrick, Mattias and Leng, Zhaoqi and Premkumar, Anjali and Bryon, Jacob and Vrajitoarea, Andrei and Sussman, Sara and others},
  title = {New material platform for superconducting transmon qubits with coherence times exceeding 0.3 milliseconds},
  journal = {Nat. Commun.},
  volume = {12},
  number = {1},
  pages = {1779},
  year = {2021},
  doi = {10.1038/s41467-021-22030-5},
  url = {https://doi.org/10.1038/s41467-021-22030-5},
}

@article{sung2021realization,
  author = {Sung, Youngkyu and Ding, Leon and Braum{\"u}ller, Jochen and Veps{\"a}l{\"a}inen, Antti and Kannan, Bharath and Kjaergaard, Morten and Greene, Ami and Samach, Gabriel O and McNally, Chris and Kim, David and others},
  title = {Realization of high-fidelity {CZ} and {ZZ}-free iSWAP gates with a tunable coupler},
  journal = {Phys. Rev. X},
  volume = {11},
  number = {2},
  pages = {021058},
  year = {2021},
  doi = {10.1103/PhysRevX.11.021058},
  url = {https://doi.org/10.1103/PhysRevX.11.021058},
}

@article{rower2024suppressing,
  author = {Rower, David A and Ding, Leon and Zhang, Helin and Hays, Max and An, Junyoung and Harrington, Patrick M and Rosen, Ilan T and Gertler, Jeffrey M and Hazard, Thomas M and Niedzielski, Bethany M and others},
  title = {Suppressing counter-rotating errors for fast single-qubit gates with fluxonium},
  journal = {PRX Quantum},
  volume = {5},
  number = {4},
  pages = {040342},
  year = {2024},
  doi = {10.1103/PRXQuantum.5.040342},
  url = {https://doi.org/10.1103/PRXQuantum.5.040342},
}

@article{zhan2026scalable,
  author = {Zhan, Ze and Li, Zishuo and Wang, Fei and Lan, Wangwei and Pan, Xianchuang and Xiang, Liang and Dou, Xu and Gao, Ran and Gong, Guicheng and Guo, Yanbo and others},
  title = {Scalable {Fluxonium} Quantum Processors via Tunable-Coupler Architecture},
  journal = {arXiv preprint arXiv:2604.13363},
  year = {2026},
  eprint = {2604.13363},
  archivePrefix = {arXiv},
  url = {https://arxiv.org/abs/2604.13363},
}

@article{krantz2019quantum,
  author = {Krantz, Philip and Kjaergaard, Morten and Yan, Fei and Orlando, Terry P and Gustavsson, Simon and Oliver, William D},
  title = {A quantum engineer's guide to superconducting qubits},
  journal = {Appl. Phys. Rev.},
  volume = {6},
  number = {2},
  year = {2019},
  doi = {10.1063/1.5089550},
  url = {https://doi.org/10.1063/1.5089550},
}

@article{manenti2021full,
  author = {Manenti, Riccardo and Sete, Eyob A and Chen, Angela Q and Kulshreshtha, Shobhan and Yeh, Jen-Hao and Oruc, Feyza and Bestwick, Andrew and Field, Mark and Jackson, Keith and Poletto, Stefano},
  title = {Full control of superconducting qubits with combined on-chip microwave and flux lines},
  journal = {Appl. Phys. Lett.},
  volume = {119},
  number = {14},
  year = {2021},
  doi = {10.1063/5.0065517},
  url = {https://doi.org/10.1063/5.0065517},
}

@article{wang2024efficient,
  author = {Wang, Tenghui and Wu, Feng and Wang, Fei and Ma, Xizheng and Zhang, Gengyan and Chen, Jianjun and Deng, Hao and Gao, Ran and Hu, Ruizi and Ma, Lu and others},
  title = {Efficient initialization of fluxonium qubits based on auxiliary energy levels},
  journal = {Phys. Rev. Lett.},
  volume = {132},
  number = {23},
  pages = {230601},
  year = {2024},
  doi = {10.1103/PhysRevLett.132.230601},
  url = {https://doi.org/10.1103/PhysRevLett.132.230601},
}

@article{xia2025fast,
  author = {Xia, Mingkang and Zhou, Chao and Liu, Chenxu and Patel, Param and Cao, Xi and Lu, Pinlei and Mesits, Boris and Mucci, Maria and Gorski, David and Pekker, David and others},
  title = {Fast superconducting qubit control with subharmonic drives},
  journal = {Nat. Commun.},
  year = {2025},
  doi = {10.1038/s41467-025-67766-6},
  url = {https://doi.org/10.1038/s41467-025-67766-6},
}

@article{schirk2025subharmonic,
  author = {Schirk, Johannes and Wallner, Florian and Huang, Longxiang and Tsitsilin, Ivan and Bruckmoser, Niklas and Koch, Leon and Bunch, David and Glaser, Niklas J and Huber, Gerhard BP and Knudsen, Martin and others},
  title = {Subharmonic Control of a {Fluxonium} Qubit via a Purcell-Protected Flux Line},
  journal = {PRX Quantum},
  volume = {6},
  number = {3},
  pages = {030315},
  year = {2025},
  doi = {10.1103/yx15-jyl7},
  url = {https://doi.org/10.1103/yx15-jyl7},
}

@article{krinner2019engineering,
  author = {Krinner, Sebastian and Storz, Simon and Kurpiers, Philipp and Magnard, Paul and Heinsoo, Johannes and Keller, Raphael and Luetolf, Janis and Eichler, Christopher and Wallraff, Andreas},
  title = {Engineering cryogenic setups for 100-qubit scale superconducting circuit systems},
  journal = {EPJ Quantum Technol.},
  volume = {6},
  number = {1},
  pages = {2},
  year = {2019},
  doi = {10.1140/epjqt/s40507-019-0072-0},
  url = {https://doi.org/10.1140/epjqt/s40507-019-0072-0},
}

@article{you2019circuit,
  author = {You, Xinyuan and Sauls, James A and Koch, Jens},
  title = {Circuit quantization in the presence of time-dependent external flux},
  journal = {Phys. Rev. B},
  volume = {99},
  number = {17},
  pages = {174512},
  year = {2019},
  doi = {10.1103/PhysRevB.99.174512},
  url = {https://doi.org/10.1103/PhysRevB.99.174512},
}

@article{SetePurcell2014,
  author = {Sete, Eyob A. and Gambetta, Jay M. and Korotkov, Alexander N.},
  title = {Purcell effect with microwave drive: Suppression of qubit relaxation rate},
  journal = {Phys. Rev. B},
  volume = {89},
  number = {10},
  pages = {104516},
  year = {2014},
  doi = {10.1103/PhysRevB.89.104516},
  url = {https://link.aps.org/doi/10.1103/PhysRevB.89.104516},
}

@article{yang2022fpga,
  author = {Yang, Yuchen and Shen, Zhongtao and Zhu, Xing and Wang, Ziqi and Zhang, Gengyan and Zhou, Jingwei and Jiang, Xun and Deng, Chunqing and Liu, Shubin},
  title = {{FPGA}-based electronic system for the control and readout of superconducting quantum processors},
  journal = {Rev. Sci. Instrum.},
  volume = {93},
  number = {7},
  year = {2022},
  doi = {10.1063/5.0085467},
  url = {https://doi.org/10.1063/5.0085467},
}

@article{rol2020time,
  author = {Rol, Michiel A and Ciorciaro, Livio and Malinowski, Filip K and Tarasinski, Brian M and Sagastizabal, Ramiro E and Bultink, Cornelis Christiaan and Salathe, Yves and Haandb{\ae}k, Niels and Sedivy, Jan and DiCarlo, Leonardo},
  title = {Time-domain characterization and correction of on-chip distortion of control pulses in a quantum processor},
  journal = {Appl. Phys. Lett.},
  volume = {116},
  number = {5},
  year = {2020},
  doi = {10.1063/1.5133894},
  url = {https://doi.org/10.1063/1.5133894},
}

@article{Serniak_2018,
  title={Hot Nonequilibrium Quasiparticles in Transmon Qubits},
  author={Serniak, K. and Hays, M. and de Lange, G. and Diamond, S. and Shankar, S. and Burkhart, L. D. and Frunzio, L. and Houzet, M. and Devoret, M. H.},
  journal={Phys. Rev. Lett.},
  volume={121},
  issue={15},
  pages={157701},
  numpages={6},
  year={2018},
  month={Oct},
  publisher={American Physical Society},
  doi={10.1103/PhysRevLett.121.157701},
  url={https://link.aps.org/doi/10.1103/PhysRevLett.121.157701}
}

@article{Ithier_2005,
  title={Decoherence in a superconducting quantum bit circuit},
  author={Ithier, G. and Collin, E. and Joyez, P. and Meeson, P. J. and Vion, D. and Esteve, D. and Chiarello, F. and Shnirman, A. and Makhlin, Y. and Schriefl, J. and Sch\"on, G.},
  journal={Phys. Rev. B},
  volume={72},
  issue={13},
  pages={134519},
  numpages={23},
  year={2005},
  month={Oct},
  publisher={American Physical Society},
  doi={10.1103/PhysRevB.72.134519},
  url={https://link.aps.org/doi/10.1103/PhysRevB.72.134519}
}

@article{Bylander_2011,
  title={Noise spectroscopy through dynamical decoupling with a superconducting flux qubit},
  author={Bylander, Jonas and Gustavsson, Simon and Yan, Fei and Yoshihara, Fumiki and Harrabi, Khalil and Fitch, George and Cory, David G. and Nakamura, Yasunobu and Tsai, Jaw-Shen and Oliver, William D.},
  journal={Nat. Phys.},
  volume={7},
  number={7},
  pages={565--570},
  year={2011},
  month={May},
  publisher={Nature Publishing Group},
  doi={10.1038/nphys1994},
  url={https://doi.org/10.1038/nphys1994}
}

@article{Martinis_2005,
  title={Decoherence in Josephson Qubits from Dielectric Loss},
  author={Martinis, John M. and Cooper, K. B. and McDermott, R. and Steffen, Matthias and Ansmann, Markus and Osborn, K. D. and Cicak, K. and Oh, Seongshik and Pappas, D. P. and Simmonds, R. W. and Yu, Clare C.},
  journal={Phys. Rev. Lett.},
  volume={95},
  issue={21},
  pages={210503},
  numpages={4},
  year={2005},
  month={Nov},
  publisher={American Physical Society},
  doi={10.1103/PhysRevLett.95.210503},
  url={https://link.aps.org/doi/10.1103/PhysRevLett.95.210503}
}

@article{Muller_2019,
  title={Towards understanding two-level-systems in amorphous solids: insights from quantum circuits},
  author={M\"uller, Clemens and Cole, Jared H and Lisenfeld, J\"urgen},
  journal={Rep. Prog. Phys.},
  volume={82},
  number={12},
  pages={124501},
  year={2019},
  month={Oct},
  publisher={IOP Publishing},
  doi={10.1088/1361-6633/ab3a7e},
  url={https://doi.org/10.1088/1361-6633/ab3a7e}
}

@article{Yoshihara_2006,
  title={Decoherence of Flux Qubits due to $1/f$ Flux Noise},
  author={Yoshihara, F. and Harrabi, K. and Niskanen, A. O. and Nakamura, Y. and Tsai, J. S.},
  journal={Phys. Rev. Lett.},
  volume={97},
  issue={16},
  pages={167001},
  numpages={4},
  year={2006},
  month={Oct},
  publisher={American Physical Society},
  doi={10.1103/PhysRevLett.97.167001},
  url={https://link.aps.org/doi/10.1103/PhysRevLett.97.167001}
}

@article{Paladino_2014,
  title={$1/f$ noise: Implications for solid-state quantum information},
  author={Paladino, E. and Galperin, Y. M. and Falci, G. and Altshuler, B. L.},
  journal={Rev. Mod. Phys.},
  volume={86},
  issue={2},
  pages={361--418},
  year={2014},
  month={Apr},
  publisher={American Physical Society},
  doi={10.1103/RevModPhys.86.361},
  url={https://link.aps.org/doi/10.1103/RevModPhys.86.361}
}

@article{yan2026characterizing,
  author = {Yan, Zhiguang and Ge, Zi-Yong and Li, Rui and Zhang, Yu-Ran and Nori, Franco and Nakamura, Yasunobu},
  title = {Characterizing many-body dynamics with projected ensembles on a superconducting quantum processor},
  journal = {Sci. Adv.},
  volume = {12},
  number = {13},
  pages = {eaeb8213},
  year = {2026},
  doi = {10.1126/sciadv.aeb8213},
  url = {https://doi.org/10.1126/sciadv.aeb8213},
}

@article{Deng2015Observation,
  author = {Deng, Chunqing and Orgiazzi, Jean-Luc and Shen, Feiruo and Ashhab, Sahel and Lupascu, Adrian},
  title = {Observation of Floquet States in a Strongly Driven Artificial Atom},
  journal = {Phys. Rev. Lett.},
  volume = {115},
  number = {13},
  pages = {133601},
  year = {2015},
  doi = {10.1103/PhysRevLett.115.133601},
  url = {https://link.aps.org/doi/10.1103/PhysRevLett.115.133601},
}

@article{Simbierowicz2024Inherent,
  author = {Simbierowicz, Slawomir and Borrelli, Massimo and Monarkha, Volodymyr and Nuutinen, Ville and Lake, Russell E.},
  title = {Inherent Thermal-Noise Problem in Addressing Qubits},
  journal = {PRX Quantum},
  volume = {5},
  number = {3},
  pages = {030302},
  year = {2024},
  doi = {10.1103/PRXQuantum.5.030302},
  url = {https://link.aps.org/doi/10.1103/PRXQuantum.5.030302},
}

\end{document}